\documentclass{article}

\PassOptionsToPackage{numbers,compress}{natbib}
\usepackage[final]{nips_2017}

\usepackage[utf8]{inputenc} % allow utf-8 input
\usepackage[T1]{fontenc}    % use 8-bit T1 fonts
\usepackage{hyperref}       % hyperlinks
\usepackage{url}            % simple URL typesetting
\usepackage{booktabs}       % professional-quality tables
\usepackage{amsfonts}       % blackboard math symbols
\usepackage{nicefrac}       % compact symbols for 1/2, etc.
\usepackage{microtype}      % microtypography

\usepackage[normalem]{ulem}
\usepackage{graphicx}

\usepackage{paralist}
\begin{document}

\title{TensorFlow-Serving:\\ Flexible, High-Performance ML Serving}

\author{
Christopher Olston\\
\texttt{olston@google.com}
\And
Noah Fiedel\\
\texttt{nfiedel@google.com}
\And
Kiril Gorovoy\\
\texttt{kgorovoy@google.com}
\And
Jeremiah Harmsen\\
\texttt{jeremiah@google.com}
\And
Li Lao\\
\texttt{llao@google.com}
\And
Fangwei Li\\
\texttt{fangweil@google.com}
\And
Vinu Rajashekhar\\
\texttt{vinuraja@google.com}
\And
Sukriti Ramesh\\
\texttt{sukritiramesh@google.com}
\And
Jordan Soyke\\
\texttt{jsoyke@google.com}
}

\maketitle

\begin{abstract}

We describe TensorFlow-Serving, a system to serve machine learning models inside
Google which is also available in the cloud and via open-source. It is extremely
flexible in terms of the types of ML platforms it supports, and ways to
integrate with systems that convey new models and updated versions from training
to serving. At the same time, the core code paths around model lookup and
inference have been carefully optimized to avoid performance pitfalls observed
in naive implementations. Google uses it in many production deployments,
including a multi-tenant model hosting service called TFS\textsuperscript{2}.

\end{abstract}

\vspace{-2mm}
\section{Introduction}

Machine learning (ML) lies at the heart of a growing number of important
problems in many fields. While there is vast literature and software devoted to
training ML models, there has been little systematic effort around deploying
trained models in production.

Indeed, until recently within Google, ML serving infrastructure has consisted
mainly of ad-hoc, non-reusable solutions. Such solutions generally start off
simple and seemingly don't necessitate a fancy or general-purpose solution,
e.g. ``just put the models in a BigTable, and write a simple server that loads
from there and handles RPC requests to the models.''

One-off serving solutions quickly accumulate complexity as they add support for
model versioning (for model updates with a rollback option) and multiple models
(for experimentation via A/B testing), which necessitate managing server RAM
carefully while avoiding availability lapses during version transitions. Then
performance problems arise as model loading causes latency spikes for other
models or versions concurrently serving, requiring careful thread management and
other techniques to keep tail latencies in check. Next, achieving high
throughput via hardware acceleration (GPUs and TPUs) requires asynchronous batch
scheduling with cross-model interleaving and more tail latency protections. As
the application matures and researchers do more experimentation with alternative
and larger models,\footnote{Model accuracy improvements are sometimes won at the
cost of model bloat, e.g. larger embeddings, more layers or more parameters per
layer.} the models may not all fit on a single server node, and a scheme must be
devised to bin-pack models onto servers and route queries accordingly. And so
on.

Over time, an application's serving system becomes a complex piece of software.
Yet application-specific APIs, assumptions and business logic baked in at the
start prevent repurposing it. As ML applications proliferate, having custom
top-to-bottom serving code for each application becomes untenable.

We set out to build a single, state-of-the-art software framework for most ML
serving workloads at Google, and in Google's Cloud-ML offering. What we've built
is called TensorFlow-Serving, for serving TensorFlow~\cite{tensorflow} and other
types\footnote{Despite the name, the core libraries are ML-platform-agnostic in
that they treat models as a black boxes (via a safe void*-like construct), and
the other layers contain very little TensorFlow-specific logic and would be
fairly easy to generalize.} of ML models. It consists of three layers: (1) a C++
library consisting of APIs and modules from which to construct an ML server, (2)
an assemblage of the library modules into a canonical server binary, and (3) a
hosted service.

The hosted service is the easiest to use: just upload your model to it and it
gets served. It also enables us to codify best practices such as validating
model quality before serving a new version, or logging inferences to catch
training/serving skew bugs~\cite{ml-test-score}. Previously, these best
practices were not widely adopted, and much effort was undertaken to persuade
each team using ML to adopt them.

Teams with specialized hardware or datacenter needs may elect to use the binary
on their own servers.

Maximum customization is possible via our library, which splits a server's
functionality into distinct modules connected via carefully designed APIs.
We've seen many ML serving variations at Google, e.g. transitioning to new
model versions in-place versus taking a server offline, serving multiple kinds
of models (TensorFlow + SomeOtherFlow) in a given server, or receiving models to
serve from different kinds of storage or data-conveyance media. Variations like
these can be realized by configuring and composing our modules in different
ways, and/or creating custom implementations of some modules. The default module
implementations supplied in the library encapsulate subtle performance
optimizations based on hard lessons e.g. around inference tail latency, such as
particular data structures for looking up models, and being mindful of which
thread performs a slow operation like freeing a big chunk of memory.

The libraries, and an instance of the binary, are
open-source.\footnote{https://github.com/tensorflow/serving} The hosted service
is available to all teams within Google, as well as to the public via Google
Cloud Platform.\footnote{https://cloud.google.com/ml-engine}

\vspace{-1mm}
\subsection{Related Work}

There is a vast amount of literature and code devoted to machine learning. Some
prominent open-source offerings include Hadoop Mahout~\cite{mahout},
scikit-learn~\cite{scikit-learn}, Spark MLlib~\cite{spark-mllib},
TensorFlow~\cite{tensorflow} and Weka~\cite{weka}. This paper focuses
specifically on serving ML models in production, which has received relatively
little systematic attention outside of application-specific approaches.

The general-purpose approaches we are aware of are Clipper~\cite{clipper},
LASER~\cite{laser} and Velox~\cite{velox}. Of these, Clipper may be the closest
effort to TensorFlow-Serving; the two systems were developed concurrently.
Clipper and TensorFlow-Serving share a focus on remaining largely agnostic to
the specific ML technology of the models being served, and have some similar
components e.g. batching. Whereas Clipper is a research system used as a vehicle
to pursue speculative ideas, TensorFlow-Serving encapsulates the state of the
art of production infrastructure as used inside Google and Google Cloud
Platform.

Lastly, there is of course a great deal of literature on web serving and other
non-ML serving scenarios, e.g.~\cite{nginx, flash, reactor, seda}, which has
informed our work. With ML serving, the objects being served (the ML models)
are: (a) comprised of logic, not just data, giving rise to isolation and other
challenges; (b) of greatly varying sizes, and in some cases hundreds of
gigabytes such that two versions cannot reside in RAM simultaneously; (c) at
times highly dynamic, with new versions emitted from training every few
minutes\footnote{The extreme form of online learning, in which models are
updated in-place while being served, is not generally practiced at Google, in
part due to challenges with debuggability and quality assurance.}; (d) greatly
helped by hardware acclerators (GPUs and TPUs~\cite{tpu}), which necessitate
carefully managed batching.

\vspace{-2mm}
\section{Library}
\label{sec:library}

The library has two parts: (1) {\em lifecycle management} modules that decide
which models to load into memory, sequence the loading/unloading of specific
versions, and offer reference-counted access to them; (2) modules that service
RPC requests to carry out {\em inference} using loaded models, with optional
cross-request batching.

\vspace{-1mm}
\subsection{Model Lifecycle Management}

\begin{figure}
\centering
\includegraphics[width=1.0\textwidth]{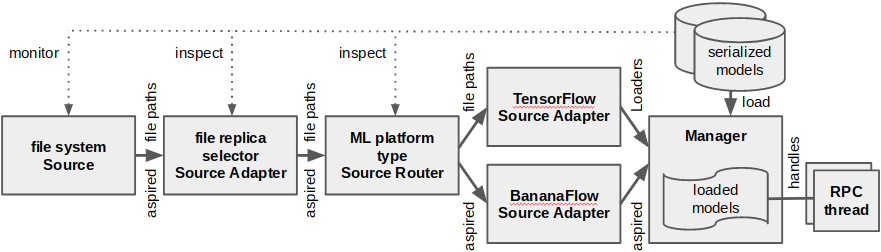}
\caption{Model lifecycle management: example topology.}
\label{fig:lifecycle}
\end{figure}

As illustrated in Figure~\ref{fig:lifecycle}, the model lifecycle management
aspect of a server consists of a chain of individual modules:
\begin{compactitem}
\item {\em Sources}, which monitor external storage systems (e.g. file systems)
to discover models and their versions.
\item {\em Source Routers} that split the stream of model versions to be loaded
based on the kind of model (e.g. TensorFlow versus BananaFlow models).
\item {\em Source Adapters}, which transform metadata about the location of each
model version into {\em Loaders} that can load the version into memory.
\item A {\em Manager} that sequences the loading and unloading of model
versions, and provides reference-counted access to them for inference.
\end{compactitem}

Other than the TensorFlow and BananaFlow Source Adapters, these modules treat
models as black boxes called {\em servables}, which could be anything. Hence the
mention of BananaFlow, a hypothetical machine learning framework unrelated to
TensorFlow. In seriousness, Google uses TensorFlow-Serving for some proprietary
non-TensorFlow machine learning frameworks as well as TensorFlow. Servables do
not need to be machine learning models at all, e.g. they could be lookup tables
that encode feature transformations.

The API used to connect these modules is called {\em aspired versions}. A call
to this API passes the name of a servable, along with a list of versions
(typically just one) that the source would like to be memory-resident.
(Implicitly, versions omitted from the list are ones it would like {\em not} to
be memory-resident.) We chose this uni-directional, idempotent API to make it
easy to build a Source that periodically polls a storage system and emits
servable versions that it aspires to reside in memory, without needing to know
which ones {\em currently} are in memory.

The aspired versions API is templated by the type of data T passed with each
version. In our canonical serving set-up, there is a Source with T = file path,
which monitors a file system and emits paths to versions of servables it desires
to reside in memory. A TensorFlow Source Adapter converts each file path string
to a TensorFlow model Loader. The Manager requires T to be Loader. Inside Google
we have production use-cases for chains of multiple Source Adapters, as well as
Source Routers and custom implementations of Sources and Source Adapters.

\vspace{-1mm}
\subsubsection{Canary and Rollback}
\label{sec:canaryAndRollback}

Our canonical file-based Source is configured with a set of servable/directory
pairs; in each directory it looks for new versions of a given servable. By
default the Source aspires the latest (largest numbered) version of each
servable, which works well for casual deployments. For production deployments,
users can override that default to address the following critical use-cases:
\begin{compactitem}
\item {\em Canary:} When a new version arrives from training and the one
      currently serving traffic becomes the second-newest version, the user can
      opt to aspire both of those versions simultaneously (i.e. load the newest
      version without unloading the older one). They would continue to send all
      prediction request traffic to the (now) second-newest version, while also
      teeing a sample of the traffic to the newest version to enable a
      comparison of their predictions. Once there is enough confidence in the
      newest version, the user would then switch to aspiring only that version
      (i.e. unload the second-newest one). This approach requires more peak
      resources, but can avoid exposing users to an important class of model
      bugs.
\item {\em Rollback:} If a flaw is detected with the current primary serving
      version (which was not caught via canary), the user can request to switch
      to aspiring a specific older version (i.e. cause the problematic version
      to be unloaded in favor of the older, presumably safe one; the order of
      the unload and load is configurable---see Section~\ref{sec:avm}). Later,
      when the problem has been addressed and a newer, safe version has been
      written from training to the file system, the user can switch to aspiring
      that one, thus ending the rollback.
\end{compactitem}

\vspace{-1mm}
\subsubsection{Aspired Versions Manager}
\label{sec:avm}

Our flagship Manager implementation is called {\em AspiredVersionsManager}. It
is parameterized by a version transition policy which is one of: (1) an
{\em availability-preserving policy} that loads a new version of a servable
before unloading the old one; (2) a {\em resource-preserving policy} that does
the opposite. The resource-preserving policy is useful for extremely large
models such that two versions cannot fit in memory at the same time, and a lapse
of availability is acceptable either because a broader system ensures there are
other replicas not currently transitioning versions, or the clients are batch
jobs that can wait/retry. We use both policies at Google.

AspiredVersionsManager incorporates several subtle performance optimizations,
some of which were discussed in~\cite{tfx}:
\begin{compactitem}
\item Read-copy-update~\cite{rcu} data structure to ensure wait-free access to
servables by inference threads.
\item Custom reference-counted servable handles that ensure the freeing of
memory for no-longer-wanted servables occurs in a manager thread, not an
inference thread. This approach avoids adding latency hiccups to inference
requests.
\item Releasing memory to the operating system upon servable unload.
\item Isolated load and inference thread pools, to insulate inference requests
from performance issues due to concurrent loading of other servables or
versions.
\item One-time use of all threads to load the initial set of servable versions,
to speed up server start-up.
\end{compactitem}
We have also made some low-level performance optimizations around the
interaction between memory allocation and CPU caches.

\vspace{-1mm}
\subsection{Inference}

We offer several TensorFlow-specific RPC APIs for inference: a low-level tensor
interface that mirrors TensorFlow's Session::Run() API, as well as higher-level
interfaces for classification and regression. To go with the latter, as well as
to integrate smoothly with training pipelines, we have co-designed a canonical
data format for examples called tf.Example. Inside Google many projects use
tf.Example with our classification or regression API, but we also have many that
use the lower-level tensor representation/API to gain more flexibility and
control over performance.

We nevertheless do our best to optimize our standard example representation
(e.g. compressing away features common to a batch of examples), and advocate for
its adoption where feasible. It offers improved type safety, but more
importantly it facilitates tools that provide other forms of safety e.g.
detecting outlier versions of a model (which can reveal training pipeline bugs)
prior to serving, and flagging training/serving skew (another bug indicator).
Such tools are part of our end-to-end ML pipeline infrastructure
(Section~\ref{sec:tfx}). To this end, we are working on additional high-level
APIs e.g. for sequence models. Our goal is to cover all but the most exotic
use-cases with typed APIs.

Each API has an RPC handler that fetches a servable handler from the Manager,
dereferences it and invokes a method such as Session::Run(), and then discards
it. The handlers are equipped with logging capability, which is useful for
debugging, detecting training/serving skew, and validating model
changes~\cite{rules}.

\vspace{-1mm}
\subsubsection{Inter-Request Batching}

As mentioned above, machine learning can leverage custom hardware e.g. GPUs and
TPUs~\cite{tpu}. The key is to combine many inference requests into a single
merged request, e.g. by concatenating the underlying input tensors. This
strategy can boost throughput substantially, but it has to be managed carefully
to avoid unduly hurting latency.

TensorFlow-Serving comes with a core library of batching primitives that is
templatized on the type of request being batched (be it tensors or some other
data). The core library supports multiple batching queues, to batch requests for
multiple servables or versions separately, and schedule them in a round-robin
fashion onto a single shared device e.g. GPU. The set of queues can be dynamic,
added and removed as servable versions come and go.

The core batching library is wrapped in two ways for use with TensorFlow: (1)
an implementation of TensorFlow's Session abstraction that batches multiple
Run() calls together, concatenating their input tensors, and then forwards to
the wrapped Session's Run(); (2) special Batch and Unbatch ops~\cite{batch-ops}
that can be inserted into a TensorFlow graph around a set of regular ops, which
pass batched data to those ops. The latter is new and not yet fully vetted, but
we are optimistic that it can supplant the former because it is more flexible.
In particular, it can be used to batch just the GPU/TPU portion of a graph,
batch the body of a sequence model's while-loop, or independently batch multiple
subgraphs e.g. the encode and decode phases of a sequence-to-sequence model. It
bears similarities to the batching approach of~\cite{on-the-fly-batching}.

\vspace{-2mm}
\section{Canonical Binary and Hosted Service}
\label{sec:service}

While our library offers great flexibility, and we have use-cases that leverage
that flexibility, many use-cases can make due with a ``vanilla'' set-up
consisting of a file-system-monitoring Source, a TensorFlow Source Adapter and
a Manager. We package this set-up as a binary so that most users do not need to
fuss with our lower-level library offering.

But our ultimate goal is to offer model serving as a hosted service, freeing
users from even running jobs. We want to raise the serving abstraction from
``run these jobs, each of which serves a set of models'' to ``serve these
models,'' with the jobs managed on their behalf. The hosted
\uline{T}ensor\uline{F}low-\uline{S}erving \uline{s}ervice is called
``TFS\textsuperscript{2}.''

\vspace{-1mm}
\subsection{TFS\textsuperscript{2}}

\begin{figure}
\centering
\includegraphics[width=0.9\textwidth]{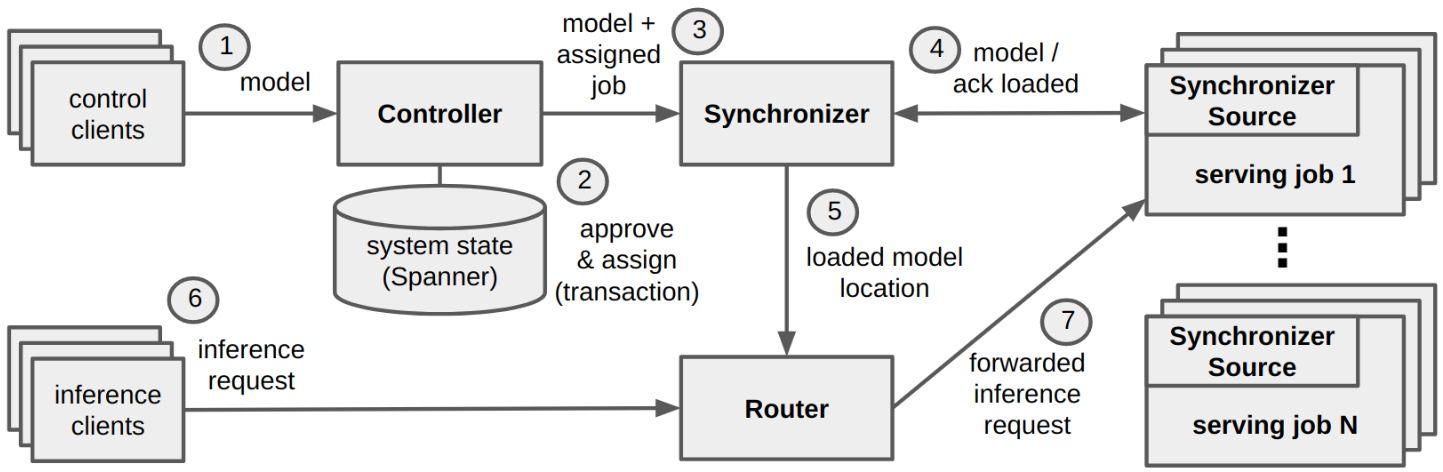}
\caption{TFS\textsuperscript{2} (hosted model serving) architecture.}
\label{fig:servomatic}
\end{figure}

In TFS\textsuperscript{2}, users issue high-level commands such as ``add model,'' ``remove
model,'' and ``add model version.'' The TFS\textsuperscript{2} infrastructure, illustrated
in Figure~\ref{fig:servomatic}, takes care of the rest, including assigning each
model to one of a suite of serving jobs based on resource fit.

The {\em Controller} takes care of adding, removing and updating users' models,
as well as honoring canary and rollback requests (see Section~\ref{sec:canaryAndRollback}).
It estimates the RAM required to serve a given model and selects a serving job
that has enough memory capacity.
Compute capacity is provisioned via user-supplied ``hints'' in advance of major production
launches; experimental launches and gradual production traffic variations are
handled automatically by a separate system that reactively auto-scales each serving
job (dynamically adding and removing job replicas as load fluctuates).
The Controller keeps all its state in Spanner~\cite{spanner}, a
globally-replicated database system, and manages it transactionally.

Once assigned a serving job by the Controller, models are disseminated to a
{\em Synchronizer} job in each data center configured to serve models.
The Synchronizer instructs serving jobs which models/versions to keep loaded at
a given time, via a special RPC-based Source library component that receives
Synchronizer instructions, and reports back status. The Synchronizer informs a
{\em Router} job which models are successfully loaded in which serving jobs, so
it can forward inference RPC requests appropriately. The Router uses hedged
backup requests~\cite{backup-requests} to mitigate latency spikes from transient
server issues or inter-request or -model interference.

We offer two TFS\textsuperscript{2} instances: (1) a Temp instance where employees taking
machine learning courses or experimenting with new types of models can try them
out, and (2) a Prod instance for robust, 24/7 serving of production traffic.
Within each instance there are several {\em partitions} which represent
specialization based on hardware (e.g. we offer partitions with TPUs) or
geography (e.g. a partition with jobs located in South America).

The serving jobs in TFS\textsuperscript{2} use the same binary\footnote{The Source to
activate---RPC-based or file-system-based---is configurable; TFS\textsuperscript{2} uses
the former while standalone jobs use the latter.} we make available for users
who wish to run their own jobs. This approach reduces the maintenance burden,
and also allows us to canary binary releases in our Temp instance before rolling
out the release more broadly to both non-hosted and hosted users.

\vspace{-1mm}
\subsection{End-to-End ML Pipelines}
\label{sec:tfx}

TensorFlow-Serving and TFS\textsuperscript{2} are part of Google's overall machine learning
infrastructure~\cite{tfx}. Other key components include model training, quality
validation (comparing inference results versus prior trained versions),
robustness validation (ensuring a model does not induce a server to crash), and
detection of training/serving skew. Google users can set up pipelines consisting
of these steps, which inject successful model versions into either stand-alone
serving jobs or TFS\textsuperscript{2}.

\vspace{-2mm}
\section{Project Status, Performance and Adoption}
\label{sec:status}

We started the project in the fall of 2015 and open sourced the library code in
winter 2016. The binary came after, and was incorporated into Google's Cloud
Machine Learning Engine offering in fall 2016. We launched the Temp instance of
TFS\textsuperscript{2} in fall 2016, and the Prod instance in winter 2017.

While always a work-in-progress, the current level of system performance is
quite good: In terms of throughput, the main bottlenecks lie in the RPC and
TensorFlow layers; we determined that TensorFlow-Serving itself can handle about
100,000 requests per second per core,\footnote{Measurements taken on a 16 vCPU
Intel Xeon E5 2.6 GHz machine.} if those two layers are factored out.
Latency-wise, we have been able to rein in tail latency substantially while
other models or versions are loading, compared to our initial naive
implementation; details and performance results are reported in~\cite{tfx}.

Each of the form-factors we offer (library, binary, service) is being used in
production inside Google, supporting various internal systems and user-facing
products. The total number of Google projects using TensorFlow-Serving is in the
hundreds. Overall traffic from Google adoption is in the tens of millions of
inferences per second. External adoption includes one-off usage e.g.
\cite{zendesk-tensorflow}, as well as projects by
Hortonworks~\cite{hortonworks-tensorflow}, IBM and SAP to incorporate
TensorFlow-serving into their general-purpose machine learning and database
platforms.

\vspace{-2mm}
\section*{Acknowledgements}

The Synchronizer system used in TFS\textsuperscript{2} as well as elsewhere in
Google, was developed by the Ads Serving team including Nels Beckman,
Andrew Bernard, Scott Daley and Ahmed Shalaby.

TFS\textsuperscript{2} is made excellent by Google's ML-SRE team: Travis Abbott,
Chris Farrar, Scott Martin, Brian McBarron, Daniel Papasian, Steven Ross and
Sean Sechrist.

We are also grateful for the ideas, contributions and collaborations from the
following individuals and teams:
Robson Araujo, Manasi Joshi, Andrey Khorlin, Alex Passos, Gautam Vasudevan, Jarek Wilkiewicz,
Google CloudML team (especially Bhupesh Chandra, Polina Dudnik and Robbie Haertel),
TensorFlow team, TFX team,
Berkeley RISELab (in particular Dan Crankshaw and Joseph Gonzalez),
and the open-source GitHub community.

\bibliographystyle{unsrtnat}
\bibliography{paper}

\end{document}